\documentclass[preprint,12pt]{elsarticle}

\usepackage{amssymb}
\usepackage{amsmath}
\usepackage{color}

\journal{NIM-A}

\begin{document}

\begin{frontmatter}



\title{High-Pressure Reinforced Vessel, produced by Additive Manufacturing, to be used as a gas target for the measurement program at n\_TOF - Multiple Argon Experiments (MArEX)}


\author[1]{D. Cortis\corref{cor1}} 
\ead{daniele.cortis@lngs.infn.it}
\author[2]{E. M. Gonzalez} %% Author name
\author[1]{D. Tatananni} %% Author name
\author[2]{P. Mastinu} %% Author name
\author[1]{Donato Orlandi} %% Author name

\cortext[cor1]{Corresponding author}

\affiliation[1]{organization={National Institute for Nuclear Physics (INFN), Gran Sasso National Laboratory (LNGS)},%Department and Organization
            addressline={via G. Acitelli 22}, 
            city={Assergi},
            postcode={67100}, 
            state={AQ},
            country={Italy}}

\affiliation[2]{organization={National Institute of Nuclear Physics (INFN), Legnaro National Laboratory (LNL)},%Department and Organization
            addressline={Viale dell'Università 2}, 
            city={Legnaro},
            postcode={35020}, 
            state={PD},
            country={Italy}}

%% Abstract
\begin{abstract}
The n\_TOF collaboration has expressed interest in developing transmission experiments with liquid argon (LAr) and gaseous argon (GAr) to improve the understanding of neutron propagation and capture in Ar. This topic is relevant to both nuclear and particle physics, as LAr is widely used in neutrino experiments, dark matter searches and other applications. Despite its widespread use, neutron interactions on Ar are still poorly characterized. In this context, the Multiple Argon Experiments (MArEX) initiative aims to address these gaps by providing detailed measurements of neutron and Ar interactions, including resolved resonance parameters and radiative capture widths, across an unprecedented energy range from 1 eV to 200 MeV.

As part of MArEX research activities, this paper presents the development of an innovative High-Pressure Reinforced Vessel (HPRV) for GAr targets, produced by Additive Manufacturing (AM) and reinforced with carbon fibre (CF). The experimental results confirmed the capability of the vessel to withstand high operating pressure without any issues and the capability of the Multi-layers Finite Element Model (MLs-FEM) to simulate and predict the stress behaviour and the operation limit of the vessel.

\end{abstract}

%% Keywords

\end{frontmatter}



\section{Introduction}

The n\_TOF collaboration \cite{nTOF} has expressed interest in developing transmission experiments with liquid argon (LAr) and gaseous argon (GAr) to improve the understanding of neutron propagation and capture in Ar. This topic is relevant to both nuclear and particle physics, as LAr is widely used in neutrino experiments (e.g., DUNE, SBND, ICARUS, MicroBooNE)\cite{DUNE,SBND,ICARUS,MicroBooNe}, dark matter searches (e.g., DarkSide)\cite{Darkside}, and other applications. Despite its widespread use, neutron interactions on Ar are still poorly characterized.

Current knowledge gaps include the neutron total cross-section minimum near 57 keV, which affects neutron transport, and the absence of measurements in the 50–100 MeV range. Data above 100 MeV are scarce and inconsistent among evaluated nuclear data libraries (i.e., ENDF/B-VIII.0), while previous studies provided only preliminary capture cross sections at eV energies and total cross sections between 20 and 70 keV \cite{ARTIE}. The Multiple Argon Experiments (MArEX) initiative \cite{MArEX} aims to address these gaps by providing detailed measurements of neutron and Ar interactions, including resolved resonance parameters and radiative capture widths, across an unprecedented energy range from 1 eV to 200 MeV. Transmission measurements using a sample-in/sample-out cycle enable determination of absolute cross-sections without reference to standards or flux normalization.

As part of MArEX research activities, this paper presents the development of an innovative High-Pressure Reinforced Vessel (HPRV) for GAr targets, produced by Additive Manufacturing (AM) with Powder Bed Fusion – Laser Based (PBF-LB) technology and reinforced with carbon fibre (CF). Metal AM has enabled the manufacturing of a custom Al-Sc alloy geometry, reinforced with CF, which is essential to prevent gas diffusion. Thanks to such technical solution, the target can withstand pressures above 400 bar, minimizing the interaction of emitted particles with the material. The use of high-pressure gas allows the production of dense GAr targets, which is mandatory when the cross-section to be measured is very low. On the other hand, emitted particles to be detected have to pass thorough the vessel with a minimum interaction. Among neutron-compatible materials, CF is probably one of best to be used: it has high strength, low destiny and low charge number. Moreover, the scattering and the capture cross-section of carbon is smooth, quite small and well-known, making it a reliable material for Monte Carlo simulations. Particularly, the small cross-section allows the minimization of the so-called "\textit{neutron sensitivity}", that is important in the experimental capture measurements,  when neutron are scattered out from the target. In fact, materials with an high scatter cross-section enhance the difficulties to discriminate the events produced by the target, due to the interaction of neutrons with the surrounding materials. In this context, the HPRV GAr target of the MArEx experiment represents the state of the art in high-density gas target design and production. Also, it is an example of low-pressure application for the X17 experiment \cite{X17} where, with the similar technical approach, a cylindrical vessel of 25 mm diameter and 85 mm length is under designing and testing to withstand a pressure of 1200 bar with 0.5 mm of Al-Sc alloy and 1.2 mm of high elastic modulus CF.

Hence, the first part of the paper describes "\textit{Vessel Design}" and the "\textit{Material Characterization}", used to implement accurate mechanical properties in numerical models. The second part presents "\textit{Multi-layers Finite Element Model}" (MLs-FEM) of the Al-Sc alloy vessel wrapped with CFs, its "\textit{Manufacturing Process}" and, finally, the "\textit{Experimental Validation}".

%% Labels are used to cross-reference an item using \ref command.

\section{Vessel Design: theoretical analysis}\label{sec1}
\begin{figure}
\centering
\includegraphics[width=.65\textwidth]{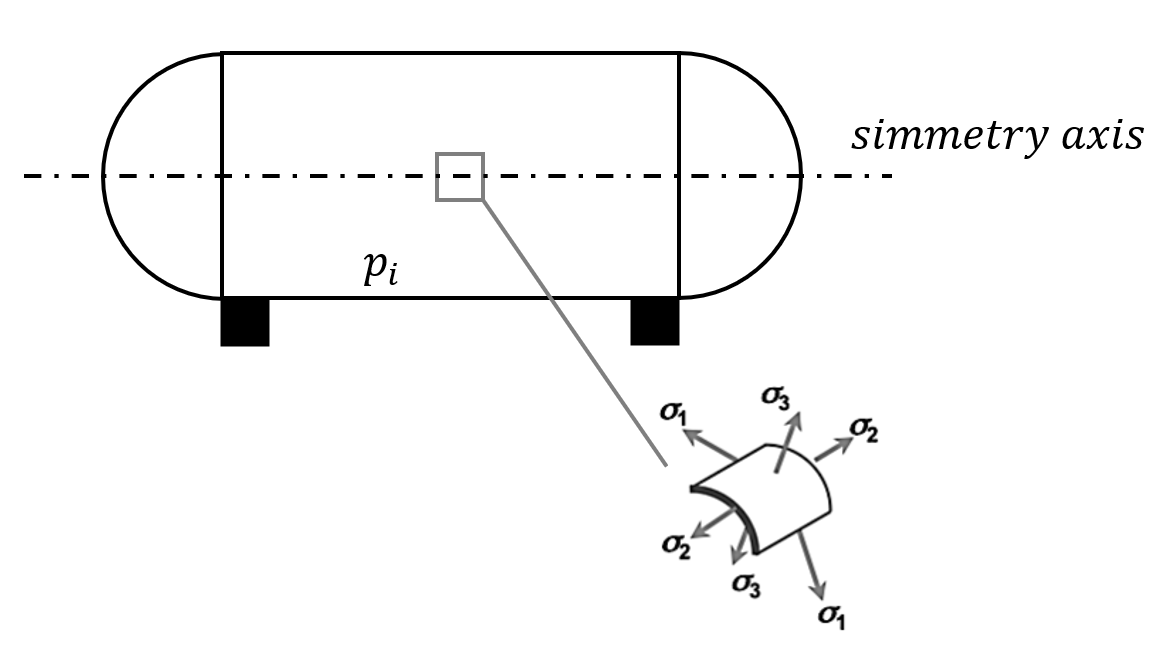}
\caption{Principal stresses direction on a generic shell of a cylindrical vessel. \label{fig:cylindrical}}
\end{figure}

Pressure vessels develop radial and tangential stresses as a function of their dimensions (i.e., radius and wall thickness). A preliminary design can be done considering some theoretical assumptions: \textit{i}) axial symmetry, \textit{ii}) small and constant wall thickness, \textit{iii}) axially symmetric loads and \textit{iv}) absence of large geometry variation. Also, referring to a cylindrical geometry, the longitudinal elongation can be considered constant around the cross-section circumference. Thanks to these assumptions, the principal stresses directions, on a generic shell of the vessel, correspond to: the tangential ($\sigma_t$ = $\sigma_1$), longitudinal ($\sigma_l$ = $\sigma_2$), and radial ($\sigma_r$ = $\sigma_3$) directions, see Figure~\ref{fig:cylindrical}. Thus, the radial and the tangential stresses along the wall thickness, in case of no external pressure, can be determined by the equations (\ref{eq:sigma_r}, \ref{eq:sigma_t}) \cite{shigley}:

\begin{equation}
\sigma_t = [(r_i^2 * p_i)/(r_o^2 - r_i^2)]*(1+r_o^2/r^2)
\label{eq:sigma_r}
\end{equation}

\begin{equation}
\sigma_t = [(r_i^2 * p_i)/(r_o^2 - r_i^2)]*(1-r_o^2/r^2)
\label{eq:sigma_t}
\end{equation}
$\\$
where, $r_i$ is the inner radius, $r_o$ the outside radius and $p_i$ the internal pressure of the vessel. Because of the internal pressure acts on the bottom of the vessel, there are also longitudinal stresses ($\sigma_l$), which can be determined by the equation (\ref{eq:sigma_l}) \cite{shigley}:

\begin{equation}
\sigma_l = [(r_i^2 * p_i)/(r_o^2 - r_i^2)]
\label{eq:sigma_l}
\end{equation}

\begin{figure}
\centering
\includegraphics[width=.70\textwidth]{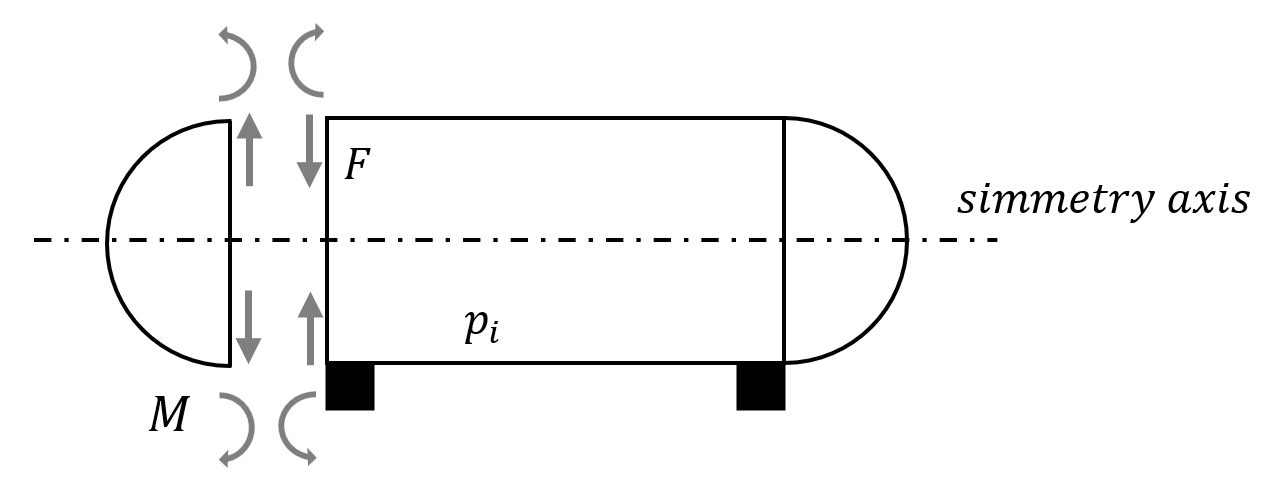}
\caption{Equilibrium of forces (F) and moments (M) for the the congruence of displacements at the ends of a cylindrical vessel. \label{fig:bottoms}}
\end{figure}

$\\$
The equations (\ref{eq:sigma_r} - \ref{eq:sigma_l}) are applicable on a cross-section of the cylinder at a significant distance from its ends and away from any areas of stress concentration. The forces and moments required to restore the congruence of displacements (Figure~\ref{fig:bottoms}) lead to an increase of the stress-state at the interface between the shell and the bottom of the vessel ($\sim$ 3 – 30\%) \cite{shigley}. As a matter of fact, in a closed cylinder, the longitudinal stress ($\sigma_l$) exists because of the pressure upon the ends of the vessel; this stress is assumed to be distributed uniformly over the wall thickness.
When the wall thickness (\textit{t}) is about one-tenth or less, of the inner radius ($r_i$/$t$ $>$ 10), the radial stress is quite small compared to the tangential one (i.e., plane stress condition) and it can be neglected ($\sigma_r=0$). Under this assumption, the tangential ($\sigma_t$) and longitudinal ($\sigma_l$) stresses can be obtained by the well-know Mariotte formulas (\ref{eq:mariotte_t}, \ref{eq:mariotte_l}) \cite{shigley}:

\begin{equation}
\sigma_t = (p_i*d_i)/2t
\label{eq:mariotte_t}
\end{equation}

\begin{equation}
\sigma_l = (p_i*d_i)/4t
\label{eq:mariotte_l}
\end{equation}
where ($d_i$) the internal diameter of the vessel. Moreover, the equivalent Von Mises stress criteria \cite{shigley} changes, as shown by the equation (\ref{eq:VM_cylinder}): 

\begin{equation}
\sigma_e = \sqrt{3}*(p_i*d_i)/4t
\label{eq:VM_cylinder}
\end{equation}

\begin{figure}
\centering
\includegraphics[width=.45\textwidth]{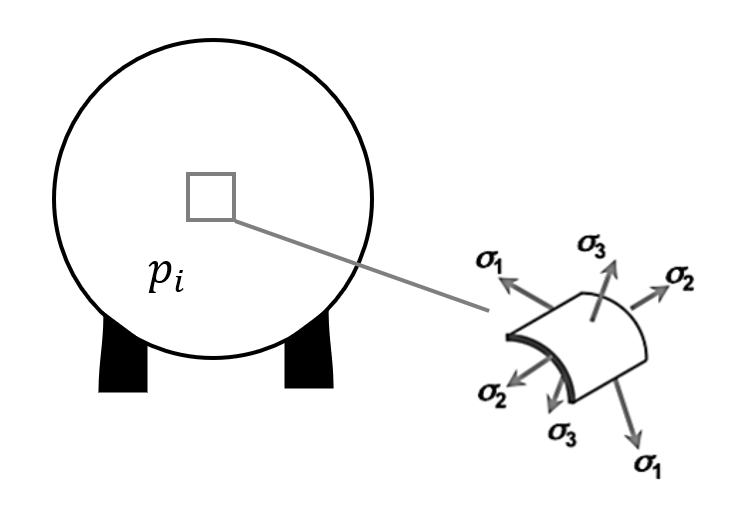}
\caption{Principal stresses direction on a generic shell of a spherical vessel. \label{fig:spherical}}
\end{figure}

$\\$
In the case of spherical vessels, the equations (\ref{eq:mariotte_t}, \ref{eq:mariotte_l}, \ref{eq:VM_cylinder}) can be further simplified. In fact, the principal stresses are equal to each other ($\sigma_t$ = $\sigma_1$ = $\sigma_l$ = $\sigma_2$) (\ref{eq:mariotte_t2}) and they have as direction any pair of orthogonal lines lying on the plane tangent to any vessel shells, see Figure \ref{fig:spherical}.

\begin{equation}
\sigma_t = \sigma_l = \sigma_e = (p_i*d_i)/4t
\label{eq:mariotte_t2}
\end{equation}

Thanks to the above considerations, a preliminary estimation of the wall thickness ($t$) of the HRPV GAr target have been done. The geometry was assumed to be spherical and, therefore, the equation (\ref{eq:mariotte_t2}) was applied. The boundary conditions, such as the Operative Pressure (OP), the Safety Factor (SF) and the Design Pressure (DP) are summarized in Table \ref{tab:boundary}. As reference material, a high-strength Al-Sc alloy was considered. The mechanical properties of this material are in the range of: Yield Strength (YS) = 420 - 440 MPa and Ultimate Tensile Strength (UTS) = 465 - 500 MPa \cite{metals4printing}. The result identified a minimum wall thickness (\textit{t}) of $\sim$ 2 mm (Table \ref{tab:results}).
However, a wall thickness of $\sim$ 2 mm would be excessive to ensure the performance of the GAr target. For this reason, it was decided to reduce the wall thickness of the Al-Sc alloy to a few tenths of  millimetre and to wrap the external surface of the sphere vessel with $\sim$ 1 - 2 mm of CF as reinforcement. Due to the complexity of the structural behaviour of such vessel solution and the difficulties of manufacturing this spherical geometry using traditional subtractive techniques, it was decided to build a Multi-layers Finite Element Model (MLs-FEM) capable of approximating the actual behaviour of the vessel and to produce the hollow sphere with metal AM by means PBF-LB technology. This choice led to the need to experimentally characterize the mechanical properties of the Al-Sc alloy.

\begin{table}
    \centering
    \begin{tabular}{ccccc} \hline 
         Operative Pressure &  $d_i$ & Safety Factor & Design Pressure \\
         (bar) &  (mm) &    & (bar) \\
          \hline 
         400 &  40 &  2 & 800 \\
          \hline 
    \end{tabular}
    \caption{HPRV GAr target design boundary conditions.}
    \label{tab:boundary}
\end{table}

\begin{table}
    \centering
    \begin{tabular}{ccccc} \hline 
         $t$ & $d_i$/t  \\
         (mm) &   \\
          \hline 
        $\sim$ 2 &  20 \\
          \hline 
    \end{tabular}
    \caption{HPRV GAr target wall thickness (\textit{t}) estimation.}
    \label{tab:results}
\end{table}

\section{Materials Characterization: Al-Sc alloy}
Mechanical properties of components produced by PBF-LB technology can change considerably depending on the process parameters and on the physical characteristics of the metal powder \cite{PBFparameters}. Hence, the Al-Sc alloy powder supplied by the Metals4Printing company (i.e., StrengthAl) \cite{metals4printing} was used to characterize the material properties and to build an elasto-plastic constitutive model to be implemented within FE simulations.

Tests were conducted at Gran Sasso National Laboratory (LNGS) of INFN (Italian Institute for Nuclear Physics) \cite{lngs} by means of a universal testing machine (i.e., INSTRON 68FM100) \cite{instron}. Tensile strains ($\epsilon$) for the determination of the YS, UTS, Elongation (A) and Elastic Modulus (E) were determined by a standard axial clip-on extensometer (i.e., INSTRON 2630-105) \cite{instron} with a gauge length of 25 mm. Tests were conducted according to EN ISO 6892-1 standard \cite{6892-1} using a constant strain rate of 0.00025 1/s (i.e., Method A222).

\begin{figure}
\centering
\includegraphics[width=.60\textwidth]{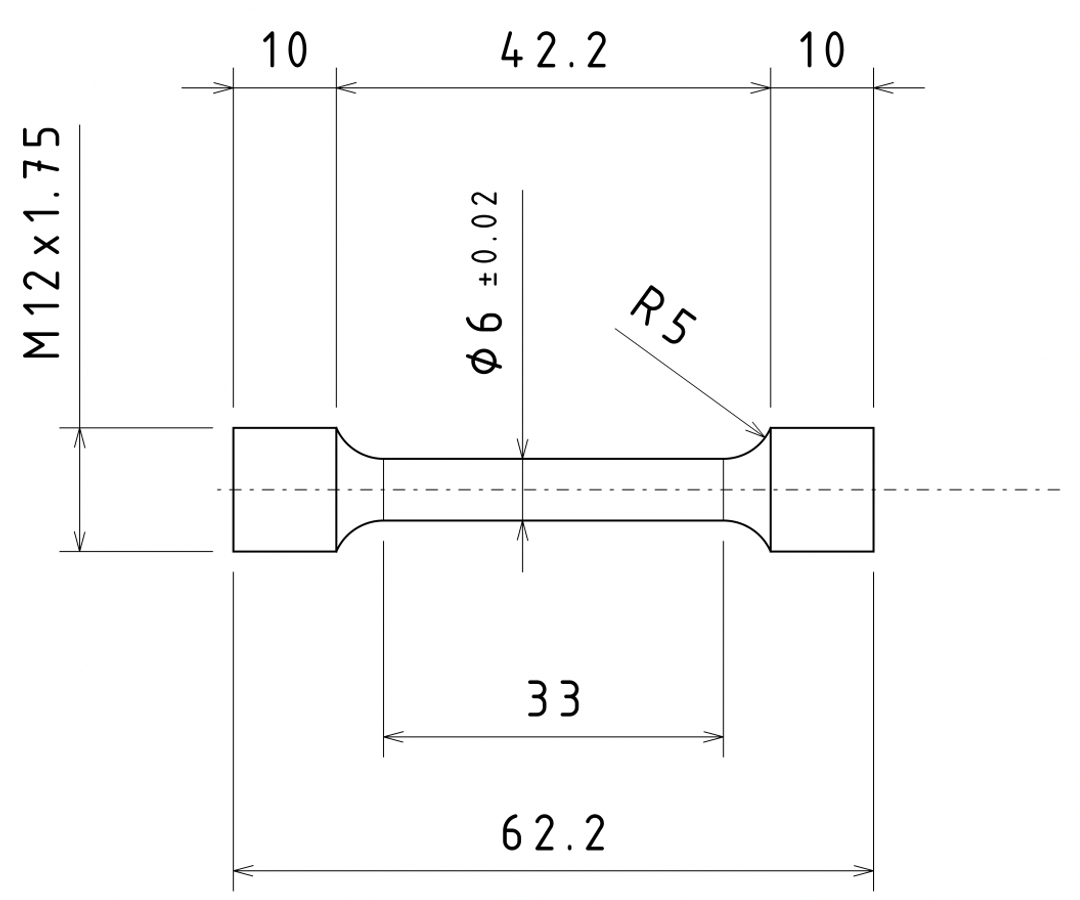}
\caption{Tensile test specimen geometry according to EN ISO 6892-1 (Annex D). \label{fig:specimen}}
\end{figure}

Tensile specimens were realized at Additive Manufacturing Lab. \cite{lngsAM} of LNGS by means a standard PBF-LB machine (i.e., PRIMA Print Sharp 150) \cite{prima} according to the EN ISO 6892-1 standard, Annex D \cite{6892-1} (Figure \ref{fig:specimen}). After the manufacturing process, the specimens were removed from the building platform by Electrical Discharge Machining (EDM) and then they were subjected to a stress relieve and a Heat Treatment (HT) of ageing (i.e., about 4h at 350 °C) \cite{scalmalloy}. The HT process has been selected to optimize and increase the mechanical strength.

\begin{figure}
\centering
\includegraphics[width=1.00\textwidth]{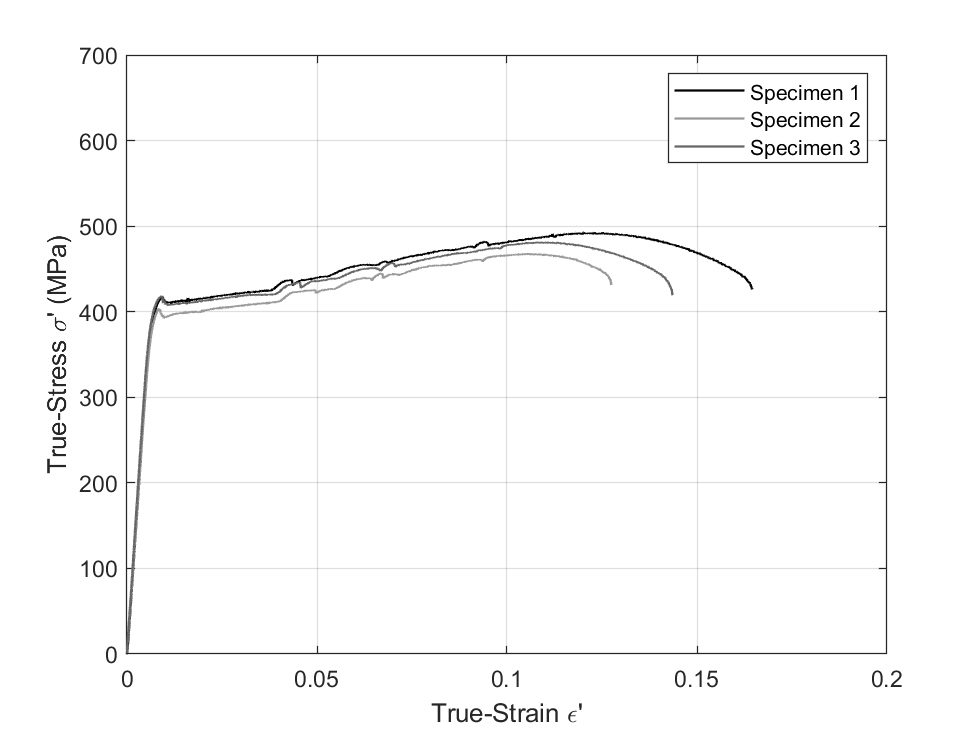}
\caption{StrengthAl: tensile tests results in terms of true-stress ($\sigma'$) and true-strain ($\epsilon'$). \label{fig:strengthAl}}
\end{figure}

The results of three tensile tests, in terms of true-stress ($\sigma'$) and true-strain ($\epsilon'$), see equation (\ref{eq:sigmatrue}, \ref{eq:epsilontrue}), are reported in Figure \ref{fig:strengthAl}, while the average mechanical properties are summarized in Table \ref{tab:strengthAl}.

\begin{equation}
\sigma' = \sigma = (1+\epsilon)
\label{eq:sigmatrue}
\end{equation}

\begin{equation}
\epsilon' = \epsilon = ln(1+\epsilon)
\label{eq:epsilontrue}
\end{equation}

\begin{table}
    \centering
    \begin{tabular}{ccccc} \hline 
         Material &  YS & UTS & E & A\\
         &  (MPa) &  (MPa) & (GPa) & (\%) \\
          \hline 
         StrengthAl &  410 &  480 & 65 & 14.5 \\
          \hline 
    \end{tabular}
    \caption{StrengthAl: average experimental mechanical properties.}
    \label{tab:strengthAl}
\end{table}

Experimental data confirm the high-strength of the Al-Sc alloy \cite{metals4printing} and the effectiveness of the PBF-LB process parameters (i.e., laser powder, scanning speed, hatch distance and layer thickness) \cite{strengthal}. The $\sigma'$-$\epsilon'$ plot shows the "\textit{Portevin–Le Chatelier}" effect \cite{portevin}, which refers to a inhomogeneous elongation due to the repeated propagation of localized deformation bands during plastic strain.

\section{Multi-layers Finite Element Model}
\begin{figure}
\centering
\includegraphics[width=.80\textwidth]{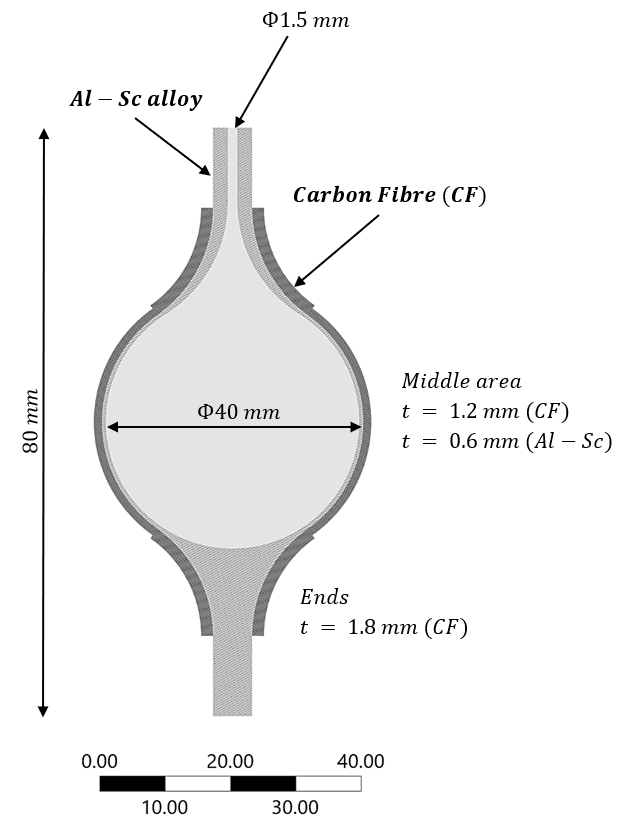}
\caption{HPRV GAr target geometry: Al-Sc alloy sphere reinforced with CF. \label{fig:vessel}}
\end{figure}

The MLs-FEM of the HPRV GAr target have been developed by means ANSYS Workbench software (2023-R2), with the combination of the Mechanical and the Composite Prep/Post (ACP) environment \cite{acp}. The Al-Sc alloy sphere were modelled with classic mechanical tools, while the CF wrapping with the ACP ones. Thanks to the ACP capability, it was possible to define the CF orientations, following the wrapping strategy on the sphere surface. The result was a 3D model composed by two solid parts bonded together with different mechanical properties based on the material and its spatial orientation, like for the CF (Figure \ref{fig:vessel}).

The middle part of the vessel was designed to minimize the wall thickness of both Al-Sc alloy and CF. In this area, the beam interacts with the gas and the presence of materials must be minimize. Thus, for the middle part of the Al-Sc alloy sphere, a wall thickness of 0.6 mm was defined, while at both ends the thickness was gradually increased to several millimetres. The transition was carried out gradually, so that stresses would not be intensified and the congruence of the displacements would be maintained. The CFs (\textit{t} = 0.3 mm for each ply) were placed on the top surface of the Al-Sc sphere with an alternating pattern (i.e., 0°/+45° respect to the symmetry axis). Similarly to the Al-Sc alloy, different thicknesses of CF were employed: 4 layers (i.e., \textit{t} = 1.2 mm) were placed in the middle part, while 6 layers (i.e., \textit{t} = 1.8 mm) were placed at both ends.

\begin{figure}
\centering
\includegraphics[width=1.00\textwidth]{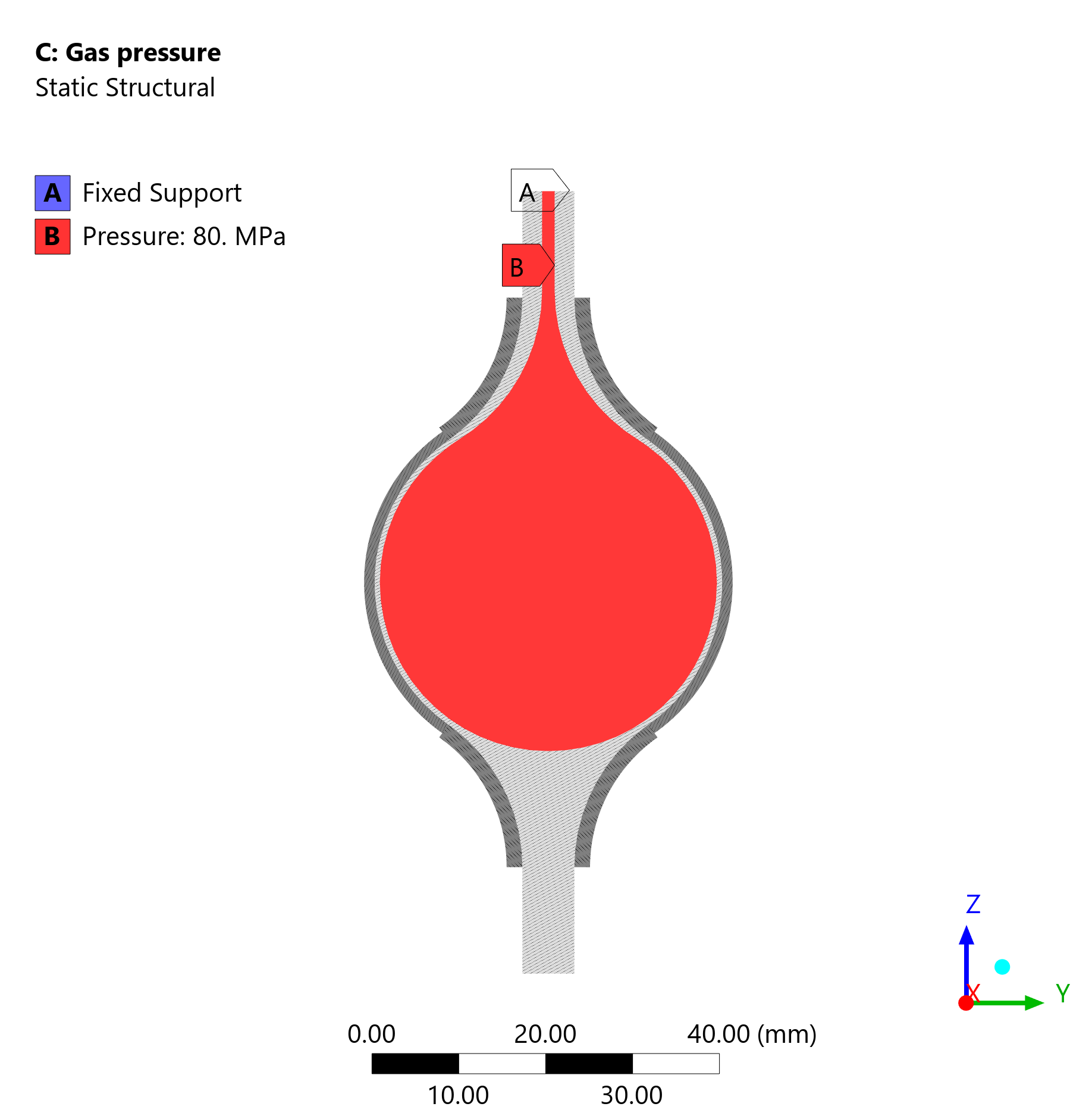}
\caption{MLs-FEM boundary conditions: inner pressure ($p_i$) and fixed support constraint.\label{fig:boundaries}}
\end{figure}

The elasto-plastic constitutive model of the Al-Sc alloy, produced by PBF-LB technology, was build using the average experimental data of Table \ref{tab:strengthAl}, while the CF properties were taken from the T1100S data-sheet (see Table \ref{tab:T1100S}) \cite{toraycma}. The MLs-FEM pressure boundary conditions (i.e., $p_i$ = 800 bar = 80 MPa) were directly applied on the inner surface of the Al-Sc sphere model, instead a fixed support constraint have been placed on the open end of the geometry. Eventually, the contact between the Al-Sc alloy sphere and the external CF was considered perfectly bonded (Figure \ref{fig:boundaries}).

\begin{table}
    \centering
    \begin{tabular}{ccccc} \hline 
         Material &  0° Tensile modulus & 0° Tensile strength & 0° Elongation \\
         &  (GPa) & (MPa) & (\%) \\
          \hline 
         T1100S &  78.8 &  1602 & 1.9  \\
          \hline  
    \end{tabular}
    \caption{T1100S carbon fibre: main mechanical properties along 0° direction.}
    \label{tab:T1100S}
\end{table}

\begin{figure}
\centering
\includegraphics[width=1.00\textwidth]{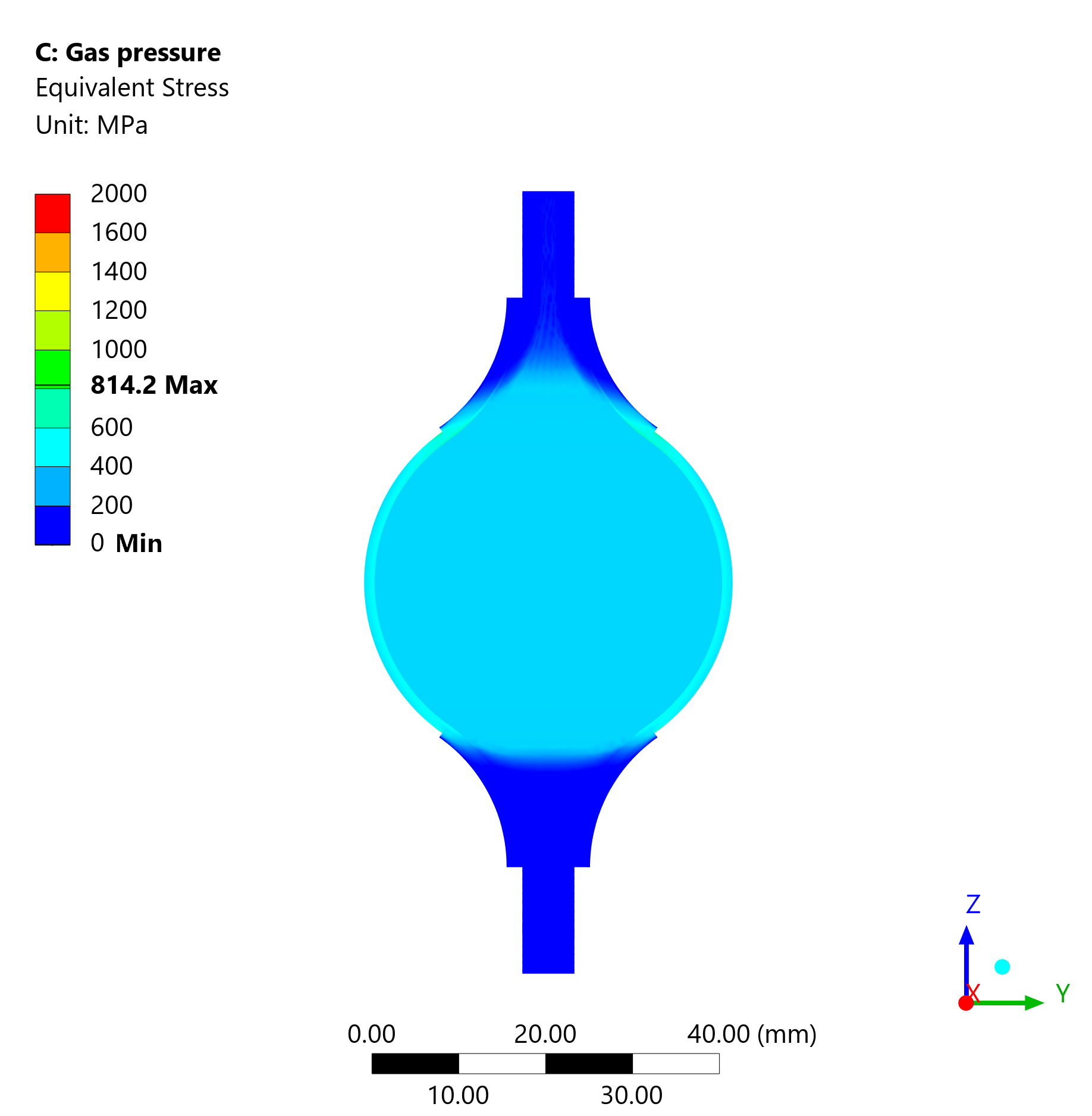}
\caption{MLs-FEM results: Equivalent Von Mises stress ($\sigma_e$).\label{fig:stress}}
\end{figure}

\begin{figure}
\centering
\includegraphics[width=1.00\textwidth]{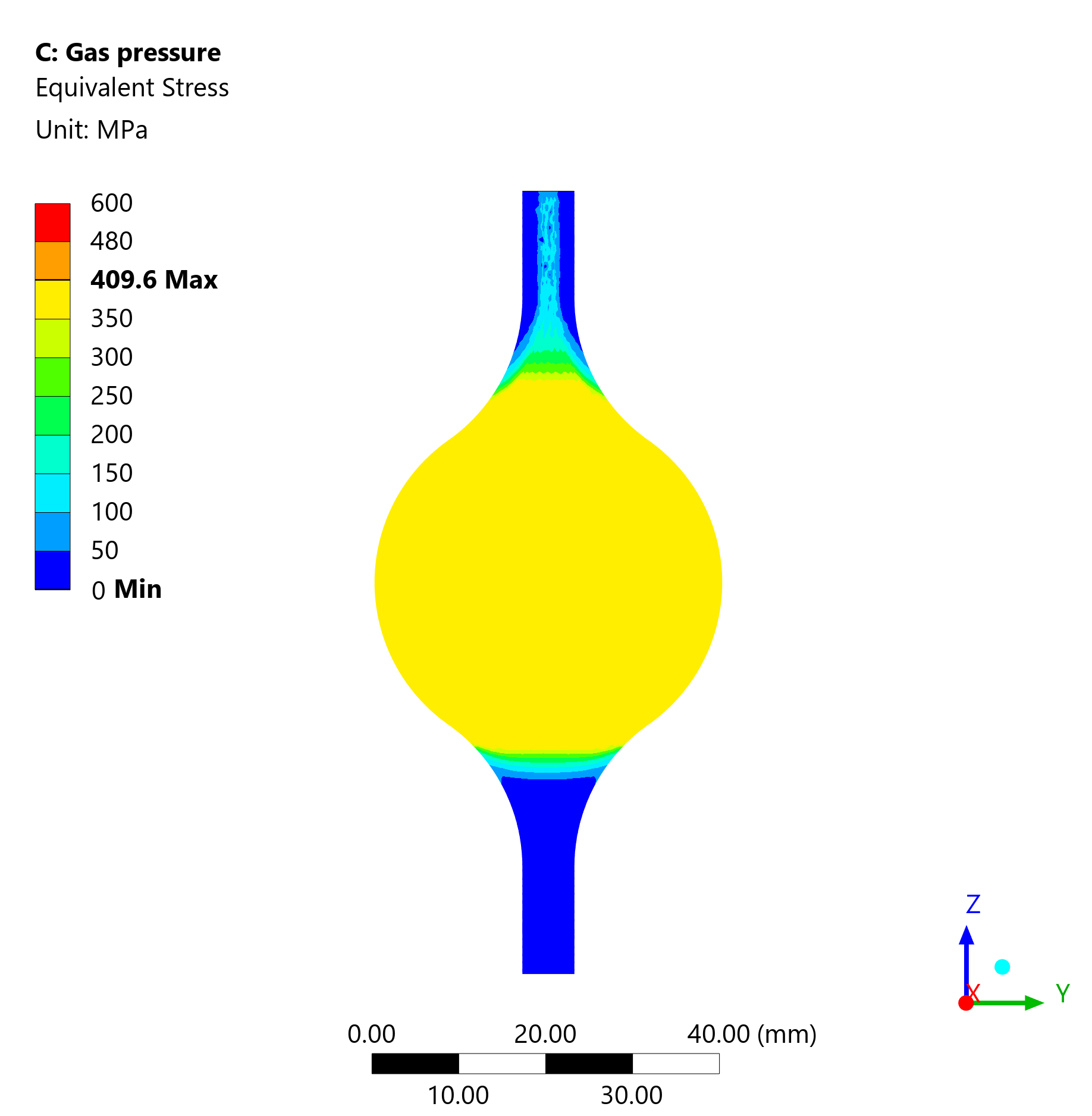}
\caption{MLs-FEM results: Equivalent Von Mises stress ($\sigma_e$) on the Al-Sc alloy core.\label{fig:stressAl2}}
\end{figure}

\begin{figure}
\centering
\includegraphics[width=1.00\textwidth]{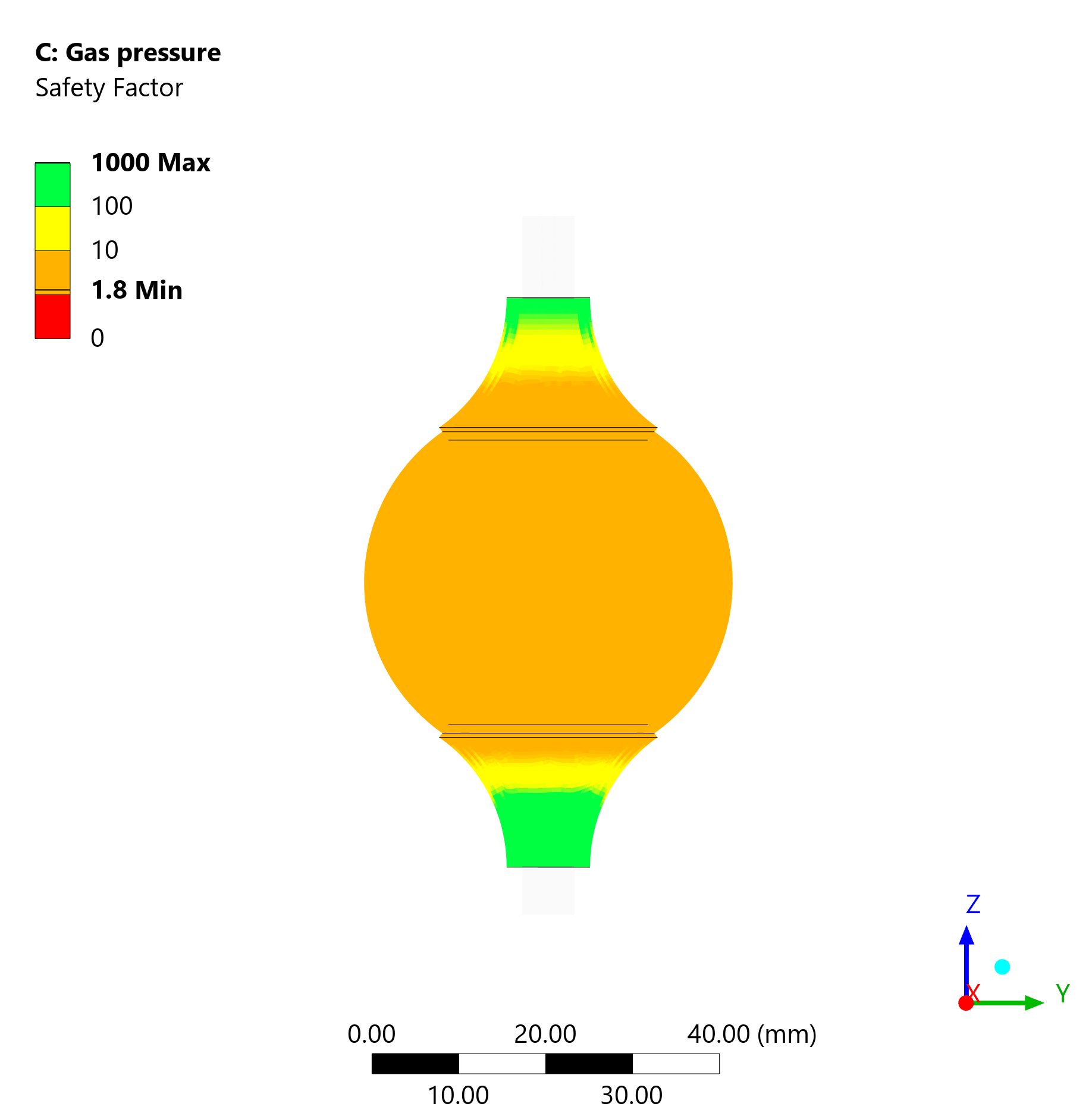}
\caption{MLs-FEM results: Safety Factor (SF) on the CF plies.\label{fig:SF}}
\end{figure}

The MLs-FEM results (Figures \ref{fig:stress} - \ref{fig:stressAl2}) show that under the DP condition of 800 bar, the HPRV GAr target achieves the incipient plasticization of Al-Sc alloy (i.e., $\sigma_e$ = $\sim$ 409.6 MPa) remaining below the UTS (Table \ref{tab:strengthAl}). In fact, most of the deformation is absorbed by the CF covering the vessel. Finally, since no information on the CF wrapping process was available at this stage of the design, the maximum stress failure criteria was applied to evaluate the SF of the plies. Under these assumptions, the SF is well above one (Figure \ref{fig:SF}), ensuring the proper operation of the vessel. In any case, a campaign of experimental tests was required to validate the design and simulations.

\section{Manufacturing Process}

\begin{figure}
\centering
\includegraphics[width=0.90\textwidth, angle =0]{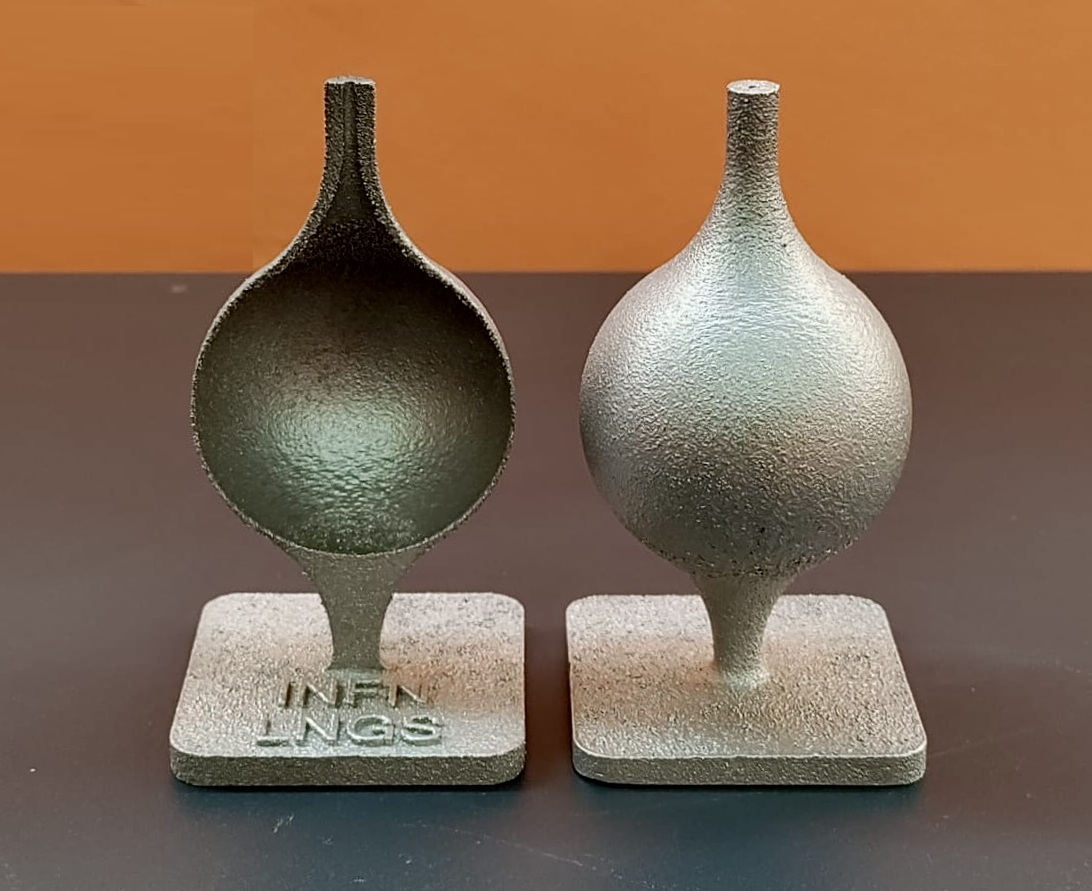}
\caption{Example of the HPRV GAr target prototype and its section after the PBF-LB production.\\ \label{fig:prototype}}
\end{figure}

HPRV GAr target prototypes have been produced at Additive Manufacturing Lab. \cite{lngsAM} of LNGS, as the tensile specimens, by means the same PBF-LB machine (i.e., PRIMA Print Sharp 150) \cite{prima}. The PBF-LB process parameters, reported in Table \ref{tab:PBFparameters}, ensure a material density $>$99\% respect to the reference of 2.67 $g/cm^3$ \cite{strengthal}. Prototypes were removed from the building platform by EDM and were subjected to the same HT of the tensile specimens. In order to have a rough surface that can ensure a good wrapping of the CFs to the Al-Sc alloy surface, no other post-operations were done, such as surface finishing. Figure \ref{fig:prototype} shows an example of the HRPV GAr target prototype and it section after the PBF-LB production.

\begin{table}
    \centering
    \begin{tabular}{ccccc} \hline 
         Laser Power &  Scanning Speed & Hatch Distance & Layer Thickness \\
         (W) &  (mm/s) & (mm) & (mm) \\
          \hline 
         170 &  715 &  0.07 & 0.04  \\
          \hline  
    \end{tabular}
    \caption{PBF-LB process parameters for Al-Sc alloy (M4P StrengthAl).}
    \label{tab:PBFparameters}
\end{table}

For each prototypes, the CF reinforcement was done by Compostiex company. \cite{compositex}. The vessels were wrapped with T1100S \cite{toraycma} using the pre-impregnated (prepreg) technique. Prepreg consists of CF fabrics pre-impregnated with an epoxy resin in which the resin and the hardener are mixed together and are activated during the manufacturing, ensuring an optimal and a uniform resin content in the final composite. This production method is considered to be one of the most advanced methods to manufacture high-performance composites. It allows a precise fibres placement, an excellent surface quality and superior mechanical properties. During the wrapping, the layers of T1100S \cite{toraycma} were applied to the HPRV GAr surface and then cured under controlled temperature and pressure in autoclave, guaranteeing the proper consolidation and structural integrity.

\section{Experimental Validation}

The experimental validation of the HPRV GAr target was carried out at the National Laboratory of Legnaro (LNL–INFN) \cite{lnl}. Defining an OP of 300 bar and applying a SF of 2, the DP during the experimental validation was set to 600 bar. The SF was defined by the CERN safety group, taking into account the compact dimensions of the target and the results of the MLs-FEM.

A total of 7 targets were wrapped by Compositex company \cite{compositex} and 5 of them underwent destructive qualification tests using a manual high-pressure oil pump capable of reaching 1500 bar. During these tests, the internal pressure was gradually increased until mechanical failure occurred. The results are reported in Table~\ref{tab:expl}. Since all tested targets exceeded the required minimum rupture threshold of 600 bar, the CERN safety group authorized their use in the experimental area. The remaining 2 targets were tested with GAr at 200 bar: one was selected for the MArEX measurement, while the second was kept as a spare.

\begin{table}
    \centering
    \caption{Experimental results of the destructive qualification tests for 5 of the HPRV GAr target wrapped by Compositex company.}
    \begin{tabular}{c c}
    \hline
       Target n.  & Failure pressure (bar) \\ \hline
        1 & \(\simeq\)800 \\ \hline
        2 & \(\simeq\)700 \\ \hline
        3 & \(\simeq\)800 \\ \hline
        4 & \(\simeq\)800  \\ \hline
        5 & \(\simeq\)700 \\ \hline
    \end{tabular}
    \label{tab:expl}
\end{table}

\begin{figure}
    \centering
    \includegraphics[width=0.8\linewidth]{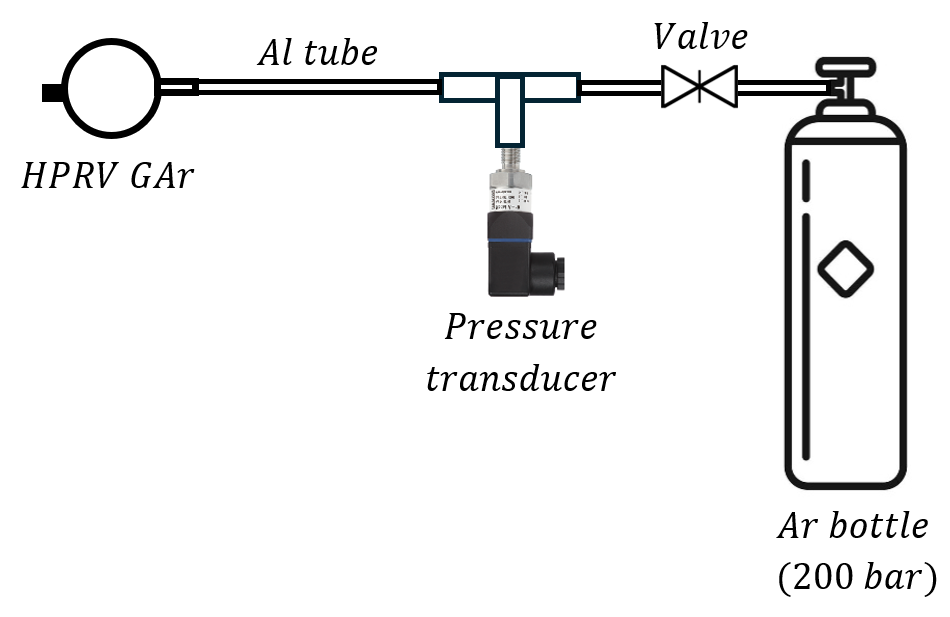}
    \caption{Layout of the filling system between the HPRV GAr target and the Ar bottle.}
    \label{fig:filled}
\end{figure}

\begin{figure}
    \centering
    \includegraphics[width=0.9\linewidth]{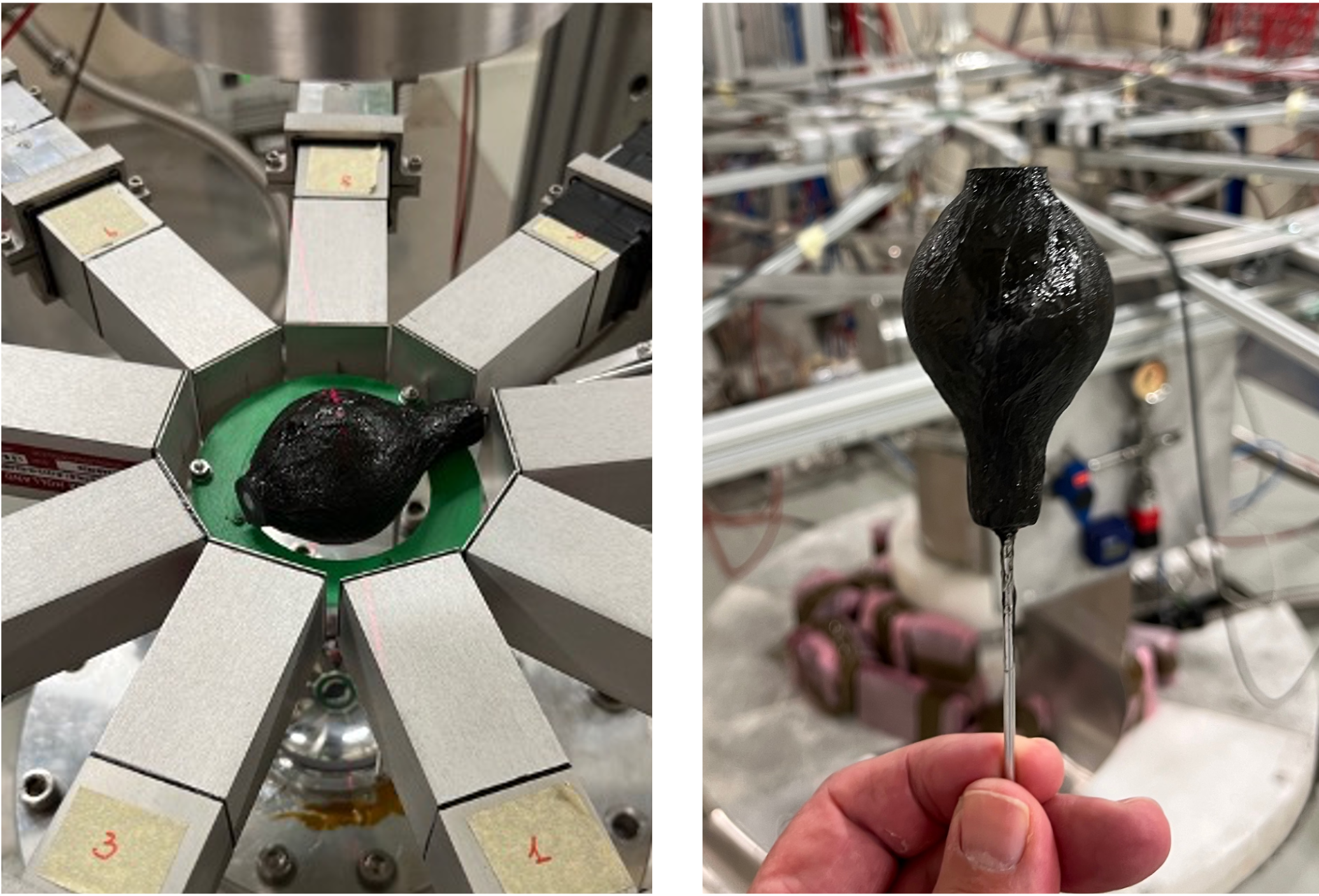}
    \caption{Final HPRV GAr target placed in the EAR2 experimental area for the MArEX measurement.}
    \label{fig:target}
\end{figure}

The configuration of the filling gas system employed for the MArEX measurement target is shown in Figure~\ref{fig:filled}. The vessel was filled with GAr at 200 bar by means a commercial pressurized bottle. A 1.5 m long aluminium tube (i.e., 0.5 mm × 2.5 mm diameter) welded to the end-cap and able to reach pressures above 1500 bar, ensured a safe transfer of the gas. A Wika pressure transducer continuously monitored the pressure. Once the pressure reached the 200 bar, the valve was closed and the vessel was disconnected from the GAr bottle and transported to the CERN experimental area. The measurements were carried out in both EAR1 and EAR2 area of the n\_TOF facility. Figure~\ref{fig:target} shows the HPRV GAr target installed along the EAR2 beam-line for the first MArEX capture campaign. During the run, the internal pressure was continuously monitored by the Wika transducer. The transmission and capture measurements were successfully completed using the HPRV GAr target. Data analysis of both experiments is currently underway and the corresponding results will be presented in a forthcoming publication.

%Safety against overpressure is assured by the rupture disc, calibrated at 350 bar. The rupture disk a metal disk, commercially available and certified, calibrated to be break at the calibrated pressure (350 bar in our case). In that way if there is an increase of the pressure inside the target, the disc will break, and the gas will go outside the system. In any case, a pressure transducer is going to measure continuously in real time the pressure inside the target.

\section{Conclusions}
This paper presents the development of an innovative HPRV, produced by AM with PBF-LB technology and reinforced with CF, to be used as a GAr target for the measurement program at n\_TOF - MArEX. The study consisted of a theoretical analysis for the preliminary design of the vessel and an experimental characterization for the materials and the manufacturing process. The characterization of the Al-Sc alloy, employed for the HPRV production, helped to build a reliable MLs-FEM able to simulate and predict the stress behaviour and the operation limit of the vessel. The validation was carried out, under real operating conditions, directly at both EAR1 and EAR2 of the n\_TOF facility at CERN. The obtained results, both in the characterization and validation phases, confirmed the capability of HPRV to withstand high operating pressure without any issues.

The innovative production process, which combined different technologies (i.e., PBF-LB and CF reinforcement), made it possible to realize a gas target that could not have been produce in any other way. Moreover, this is an example of a new integrated design approach, where the experimental performance of materials, due to the production process (i.e., PBF-LB), is taken into account directly into simulations to create a FEM that can be extended to different experimental situations, drastically reducing failures and implementation times. Finally, the results of this research have paved the way for other new technical and experimental solutions that can be implemented by the authors for future developments, even in different fields.

%% Refer following link for more details.
%% https://en.wikibooks.org/wiki/LaTeX/Mathematics
%% https://en.wikibooks.org/wiki/LaTeX/Advanced_Mathematics


\begin{thebibliography}{00}

%% For numbered reference style



\bibitem{nTOF}
  n\_TOF. 
  \textit{The neutron time-of-flight facility at CERN},
  https://ntof-exp.web.cern.ch.

\bibitem{DUNE}
  DUNE. 
  \textit{Deep Underground Neutrino Experiment},
  https://www.dunescience.org

\bibitem{SBND}
  SBND. 
  \textit{Short-Baseline Near Detector},
  https://sbn-nd.fnal.gov

\bibitem{ICARUS}
  ICARUS. 
  \textit{Imaging Cosmic And Rare Underground Signals},
  https://icarus.fnal.gov

\bibitem{MicroBooNe}
  MicroBooNe. 
  \textit{Micro Booster Neutrino Experiment},
  https://microboone.fnal.gov

\bibitem{Darkside}
  DarkSide. 
  \textit{Dark Matter Experiment},
  https://darkside.lngs.infn.it

\bibitem{ARTIE}
  Andringa, S., et al.,
  \textit{Measurement of the total neutron cross section on argon in the 20 to 70 kev energy range},
  Phys. Rev. C 108,
  L011601 (2023),
  https://doi.org/10.1103/PhysRevC.108.L011601.

\bibitem{MArEX}
  Andringa, S., et al.,
  \textit{Multiple Argon Experiments at nTOF (the MArEX initiative)},
  CERN-INTC-2023-046,
  INTC-I-256 (2023),
  https://cds.cern.ch/record/2856506/files/INTC-I-256.pdf

\bibitem{X17}
  Gustavino, C., et al.,
  \textit{Searching for the X17 Particle using the novel n+3He Reaction},
  Proposal to the ISOLDE and Neutron Time-of-Flight Committee,
  (2025),
  hhttps://cds.cern.ch/record/2920978/files/INTC-P-727.pdf

\bibitem{shigley}
  Richard G. Budynas, J. Keith Nisbett,
  \textit{Shigley’s Mechanical Engineering Design},
  10th edition,
  MC GrawHill,
  2014.

\bibitem{metals4printing}
  Metals4Printing (M4P) Material Solution GmbH – Austria. 
  \textit{StrengthAl - Technical Data Sheet - Rev.V05/1-20},
  https://www.metals4printing.com.

\bibitem{PBFparameters}
  Cortis, D., Pilone, D., Broggiato, G. et al.,
  \textit{Setting of L-PBF parameters for obtaining high density and mechanical performance of AISI 316L and 16MnCr5 alloys with fine laser spot size.},
  Prog Addit Manuf 9,
  2017–2029 (2024),
  https://doi.org/10.1007/s40964-023-00556-y.

\bibitem{lngs}
  Gran Sasso National Laboratory, 
  https://www.lngs.infn.it/en

\bibitem{instron}
  INSTRON Company, 
  https://www.instron.com/

\bibitem{prima}
  PRIMA ADDITIVE, 
  https://www.primaadditive.com/
  
\bibitem{6892-1}
  EN ISO 6892-1:2020. 
  \textit{Metallic materials — Tensile testing — Part 1: Method of test at room temperature}.

\bibitem{strengthal}
  Cortis, D., Pilone, D., Grazzi, F. et al.,
  \textit{Functionally graded material via L-PBF: characterisation of multi-material junction between steels (AISI 316L/16MnCr5), copper (CuCrZr) and aluminium alloys (Al-Sc/AlSi10Mg)},
  Prog Addit Manuf 10,
  2455–2472 (2025).,
  https://doi.org/10.1007/s40964-024-00761-3.
  
\bibitem{lngsAM}
  Orlandi, D., Cortis, D.,
  \textit{Metal additive manufacturing at INFN-LNGS laboratory: Facilities, testing and future capabilities},
  AIP Conf. Proc. 2908,
  030001 (2023),
  https://doi.org/10.1063/5.0161106.

\bibitem{scalmalloy}
  Cortis, D., Campana, F., Orlandi, D. et al.,
  \textit{Strength and fatigue behavior assessment of the SCALMALLOY® material to functionally adapt the performance of L-PBF components within CAE simulations.},
  Prog Addit Manuf 8,
  933–946 (2023).,
  https://doi.org/10.1007/s40964-022-00366-8.

\bibitem{portevin}
   Zhang P., Liu G., Sun J.,
  \textit{A critical review on the Portevin-Le Chatelier effect in aluminum alloys.},
  J Cent South Univ 29:744–766. (2022),
  https://doi.org/10.1007/s11771-022-4977-x.
  
\bibitem{acp}
  ANSYS Software. Workbench 2023-R2,
  \textit{Composite Prep/Post (ACP)},
  https://developer.ansys.com/docs/acp.

\bibitem{toraycma}
  Toraycma Company. 
  \textit{T1100S Data-Sheet},
  https://www.toraycma.com/wp-content/uploads/T1100S-Data-Sheet.pdf

\bibitem{compositex}
  Compositex srl. a HRC group company,
  https://compositex.com.

\bibitem{lnl}
  National Laboratory of Legnaro 
  https://www.lnl.infn.it/en/
  
\end{thebibliography}
\end{document}